\documentclass[prd, twocolumn, nofootinbib]{revtex4}
\usepackage{graphicx, epsfig}

\def\fun#1#2{\lower3.6pt\vbox{\baselineskip0pt\lineskip.9pt
  \ialign{$\mathsurround=0pt#1\hfil##\hfil$\crcr#2\crcr\sim\crcr}}}

\begin{document}

\title{Gravitational Lensing by Cosmic Strings in the Era
of Wide-Field Surveys}

\author{Dragan Huterer and Tanmay Vachaspati}
\affiliation{Department of Physics, Case Western Reserve University, 
Cleveland, OH~~44106}

\begin{abstract}
Motivated by the recent claim for gravitational lensing by a cosmic
string, we reinvestigate the probability of finding such an event with
upcoming wide-field surveys. If an observed lensing event is suspected
to be due to a string, observations of the vicinity of the event in a
circle of diameter $L$ centered on the observed lens should reveal
several additional lensing events. For a string located nearby
($z\lesssim 0.5$), we find that observations in a region of size
$\approx 1$ arcmin$^2$ will see $\sim 100$ objects, of which $\sim 5$
would be lensed by the string, compared to $\sim 0.1$ lensed by
conventional sources.
\end{abstract}

\maketitle

\section{Introduction}

Cosmic strings are linear sources of energy-momentum, believed to have
possibly been produced during a phase transition in the early
universe. Many cosmological signatures of cosmic strings have been
investigated over the past few decades
\cite{Vil_She_94,Hin_Kib_95}, including large-scale structure
formation, gravitational wave spectrum, effect on cosmic microwave
background anisotropies, and gravitational lensing.  Observations of
the large-scale structure and the cosmic microwave background
anisotropy show that strings are not solely responsible for
either. This places a constraint on the linear energy density (or,
tension) in a string, $\mu \lesssim 2\times 10^{22}$ gm/cm
\cite{Vil_She_94,Hin_Kib_95}, often written in the dimensionless
combination $G\mu/c^2 \lesssim 10^{-6}$ or equivalently
$\delta \equiv 8\pi G\mu/c^2 \lesssim 3\times 10^{-5}$, where $c$ is
the speed of light and $G$ is Newton's constant. Lighter strings do
not have the virtue of explaining any major cosmological
conundrum, yet are not ruled out, and would have profound implications
for particle physics and early universe cosmology.

Strings lighter than $\delta\simeq 3\times 10^{-5}$ can also be
detected since they act as lineal gravitational lenses. Indeed, since
the study of cosmic strings was first initiated, a few gravitational
lensing events have been discovered that were suspected to have been
sourced by a cosmic string \cite{Cowie_Hu,Turner_86}. However the
suspicion has not been confirmed (e.g. \cite{Hewitt}).

Our interest in string lensing was rekindled by the recent claim for a
possible lensing by a string~\cite{CSL1}, and further stoked by the
advent of deep, wide-field surveys that will revolutionize the field
of observational astrophysics.  We build on previous work
(e.g. \cite{Hog_Nar_84,Pac_86,Vil_87,Hin_90,Lai_Vac_96,Lai_Kra_Vac_97,Ber_Uza_01})
and revisit estimates of probability of lensing by a cosmic string. In
particular, we identify simple yet effective strategies for confirming
or refuting the string hypothesis, given one putative lens.

\section{Expected Number of Galaxies Lensed by a String}

Let us consider lensing by a single open cosmic string. 
We assume that the string has correlation length $\xi$
and that its length from any reference point on it, $\vec{r}$, scales
as

\begin{equation}
 l= R \left ( {R\over \xi}\right )^a
\label{eq:l_R}
\end{equation}

\noindent where $R=|\vec{r}-\vec{r_1}|$ and $l$ is the proper length of string
between $\vec{r}$ and any other point $\vec{r_1}$. The parameter $a$
is equal to one for a pure random walk of the string (so that
$R\propto\sqrt{l}$), and zero for a perfectly straight string.

Cosmic strings do not introduce spacetime curvature, but simply
produce a conical spacetime with deficit angle $\delta$
\cite{Vil_84,Gott}. Light propagation
from a source in a conical spacetime leads to gravitational lensing
with an angular separation between the images is given by~\cite{Vil_84}

\begin{equation}
\Delta\alpha= \delta\, {D_{ls}\over D_{os}}\sin{\theta}
\label{eq:splitting}
\end{equation}

\noindent where $D_{os}$ and $D_{ls}$ are angular diameter distances 
from observer to source and lens (string) to source respectively, and
$\theta \in [0,\pi ]$ is the angle between the string direction and
the line of sight\footnote{We ignore the string velocity factors in
Eq.~(\ref{eq:splitting}), since the string is expected to be at most
mildly relativistic.}. Note that the lensed object reported in
Ref.~\cite{CSL1} is located at redshift $z_s=0.46$ and its image
splitting is reported to be $\Delta\alpha=2''$; this,
Eq.~(\ref{eq:splitting}) and the bound $\delta \lesssim 3\times
10^{-5}$ imply that the putative string is located relatively close to
us: $z_l\lesssim 0.25$. [This result is roughly independent of
cosmological parameters.] The two similar images of the source in
Ref.~\cite{CSL1} also indicate that the string should be straight on
scales, $\xi \gtrsim \Delta\alpha\sim 2''$; if this were not true, the
images would be distorted as in Ref.~\cite{Lai_Kra_Vac_97}. Note that,
here and in the rest of this paper, we quote distances (such as $l$,
$R$ and $\xi$) as angles projected on the sky.

To start, we would like to calculate the number of galaxies lensed by a
single straight string at redshift $z_l$. The probability for lensing
for a single galaxy found in the survey at redshift $z_s$ due to the
infinitely long string located at $z_l$ is 

\begin{eqnarray}
P(z_l, z_s) &\simeq &{\langle \Delta\alpha\rangle \over \pi}\,
	\Theta(\Delta\alpha-\Delta\alpha_{\rm min})\\
&=& \,{{2 \delta}\over \pi^2} \,{D_{ls}\over D_{os}}\,
	\Theta(\Delta\alpha-\Delta\alpha_{\rm min}) \nonumber
\end{eqnarray}

\noindent where  $\langle\Delta\alpha\rangle$ is the angular separation 
of images averaged over string directions on the sky,
$\Delta\alpha_{\rm min}$ is the angular resolution of the survey, and
$\Theta$ is the Heaviside step function. We have assumed that the
string appears as a great circle on the sky. More realistically,
there will be many ($\sim 10$ \cite{AlbTur,AllShe,BenBou})
long strings in our horizon, each of which does not cover a
great circle on the sky. But we can expect the above estimate
to be roughly correct, the greater number of strings compensating
for the smaller extent of each of them.

\begin{figure}[!t]
\includegraphics[height=3.5in, width= 2.4in, angle=-90]{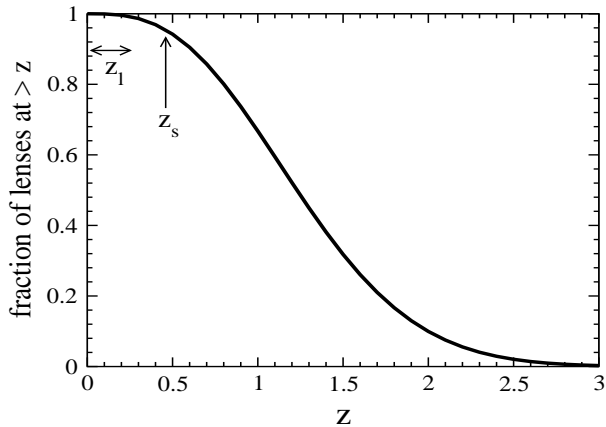}
\caption{Fraction of source galaxies {\it beyond} a given redshift $z$, assuming
the distribution listed in the text and expected in upcoming
wide-field surveys. The vertical arrow denotes the location of the
lensed source reported in Ref.~\cite{CSL1}, while the horizontal bar
denotes the range of redshifts allowed for the lens (string) in this
case. Note that the vast majority of sources are expected to be at
redshifts higher than the reported string redshift $z_l$. }
\label{fig:N_obs}
\end{figure}

The number of lenses expected is simply

\begin{equation}
N(z_l)=\int_{z_l}^{\infty} P(z_l, z_s)\,
                     {dN_{\rm src} \over dz}(z_s)\, dz_s
\label{eq:n_snap}
\end{equation}

\noindent where $z_l$ is redshift of the string and $dN_{\rm src}/dz$ 
is the number of observable galaxies at redshift $z$ in interval $dz$.
By ``observable galaxies'' we mean those that are part of the source
population imaged by the survey.  We assume that the redshift
dependence of the number density of these galaxies is given by
$dN_{\rm src}/dz\propto z^2\,\exp(-(z/z_0)^2)$ with $z_0=1.13$ (which
corresponds to median redshift of 1.23)~\cite{Refregier}. The surface
density of galaxies peaks at $z\sim 1$. If the string is located at
$z\lesssim 0.25$, as indicated by the reported string lens
candidate~\cite{CSL1}, most ($\sim 99\%$) of galaxies along the line
of sight are located behind the string; see Fig.~\ref{fig:N_obs}.

Upcoming wide-field surveys from space, such as Supernova/Acceleration
Probe (SNAP; \cite{SNAP}) and ground, such as Large Synoptic Survey
Telescope (LSST; \cite{LSST}), will cover hundreds and tens of
thousands of square degrees, perform deep imaging to about 28th and
26th magnitude in $R$ band, and find locations, photometric redshifts
and shapes for about $10^8$ and $10^{10}$ galaxies, respectively. For
$\delta \sim 10^{-5}$, this fact and Eq.~(\ref{eq:n_snap}) imply of order
$10^2$--$10^5$ galaxies lensed by the string.  This sounds fantastic
until we realize that the number of galaxies/quasars lensed by
intervening large-scale structure will be at least two orders of
magnitude higher, since the optical depth for lensing has been
measured to be around in $10^{-3}$ in the JVAS/CLASS survey
\cite{CLASS}. Further complication will be induced by the finite
angular resolution of these surveys, as well as spurious candidates,
such as binary systems. The signal we are looking for is likely to be
dwarfed by the ``noise'' just described, and it is difficult to expect
that future wide-field surveys will either detect cosmic strings
through lensing or impose much better constraints on the string scale
$\delta$. On a positive side, it is possible that we will get lucky
with a serendipitous discovery, such as that reported in
Ref.~\cite{CSL1}, and chances for that are significantly enhanced with
future wide-field surveys.

\section{Confirming the lensing by a cosmic string hypothesis}

Let us now consider a different problem.  Given the observation of one
lensing event by a cosmic string, how likely is it that we will find
another such event, and what can we do to find it?  The fact that the
location of one lensing event is known is very helpful, since we know
that the string passed through that angular location, and the
orientation of the double image tells us the direction of passage
projected on the sky.

In the following, we assume that future wide-field surveys, such as
SNAP and LSST, will find about 100 galaxies/arcmin$^2$, which
corresponds roughly to 28th magnitude in the R color band. This
estimate is conservative; for example, around 200 galaxies/arcmin$^2$
are detectable in the Hubble Deep Field.  These galaxies will be fully
resolved and their shear will be measured for weak lensing studies.

The shape of the string is described by Eq.~(\ref{eq:l_R}). There are
two limiting cases:

\begin{figure}[!t]
\includegraphics[height=2.6in, width= 3.5in]{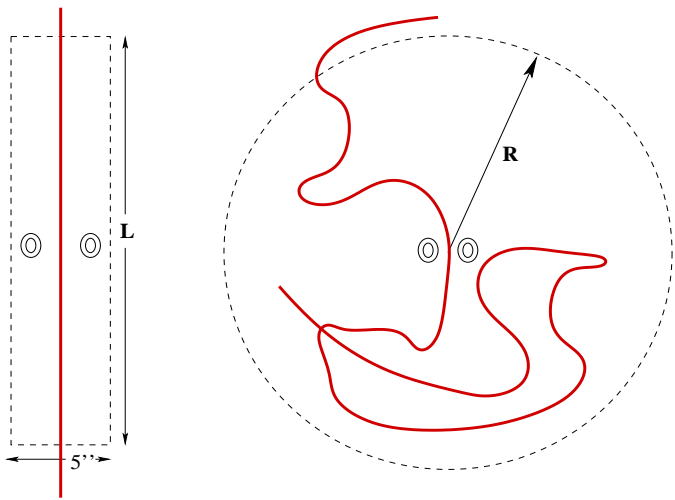}
\caption{Two extreme cases discussed in the text: straight string (left) and
string exhibiting a random walk (right). In each case the cosmic
string is shown with solid lines, while the area that we need to image
in order to find at least one more lens is enclosed with dashed
lines. The pair of concentric circles denotes images of the original lensing
event.}
\label{fig:string}
\end{figure}

\begin{figure}[!t]
\includegraphics[height=3.5in, width =2.7in, angle=-90]{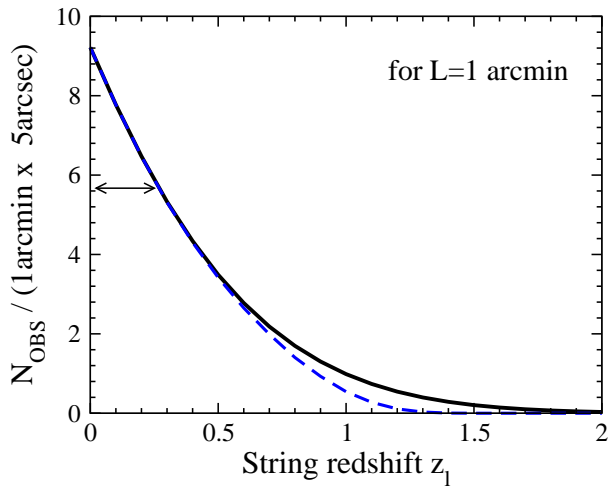}\\
\includegraphics[height=3.5in, width=2.7in, angle=-90]{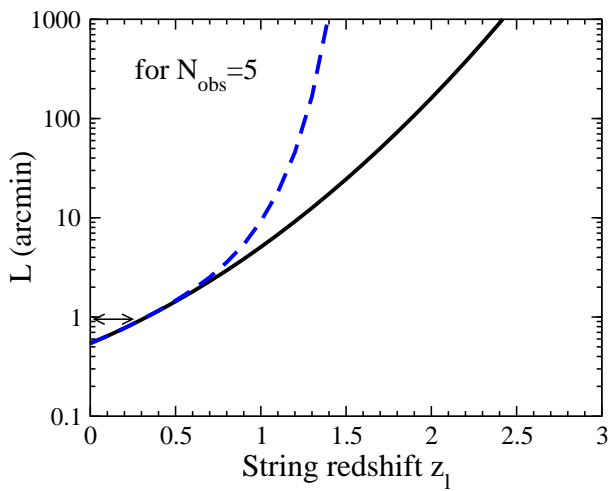}
\caption{Top panel: number of galaxies found in the area $5''\times 1'$ 
as shown in the left drawing in Fig.~\ref{fig:string} (i.e. for
$L=1'$) as a function of the redshift of the string, $z_l$. We have
assumed the surface density of galaxies of $N_s/A=100$
gal/arcmin$^2$. Solid line assumes a perfect angular resolution of the
survey, $\Delta\alpha_{\rm min}=0$, while the dashed line assumes the
resolution of $\Delta\alpha_{\rm min}=1''$. The horizontal line shows
the range of redshifts allowed for the string reported in
Ref.~\cite{CSL1}.  Bottom panel: same as above, except we now show the
required follow up length $L$ in order to see 5 additional lensing
events ($N_{\rm obs}=5$), as a function of string redshift $z_l$. Note
that the two panels contain the same information, and are both shown
for clarity.}
\label{fig:surface_dens}
\end{figure}

1) $a=0$: the string is straight, static and perpendicular to the line
of sight.  In that case an obvious strategy is to look a distance $L$
along the axis of the observed lens (see Fig.~\ref{fig:string}, left 
drawing). From Eq.~(\ref{eq:splitting}) it follows that the maximum
angular splitting for experimentally allowed values of $\delta$ is
about 5'', so a sensible strategy is to look 5'' perpendicularly and
symmetrically to the aforementioned axis; see
Fig.~\ref{fig:string}. This guarantees that all galaxies lensed by
this chunk of the string of length $L$ will be seen. The number of
lensed galaxies depends on the string redshift, $z_l$, as

\begin{eqnarray}
N_{\rm obs}(z_l) 
&=& L\, \int_{z_l}^{\infty} {d(N_{\rm src}/A)\over dz_s}\,
    	        \Delta\alpha(z_l, z_s) \nonumber \\[0.1cm]
&&		\times\, \Theta\left (\Delta\alpha(z_l, z_s)-
		\Delta\alpha_{\rm min}\right )\, dz_s  
\label{eq:N_obs}
\end{eqnarray}

\noindent where $N_{\rm src}/A$ is the surface density 
of source galaxies. The expected number of sources for $L=1'$ is shown
in the top panel of Fig.~\ref{fig:surface_dens}.  For a locally
straight string and the lens redshift $z \lesssim 0.5$, it is
sufficient to look about $L=1'$ along the axis passing between the
images.  In doing so we are guaranteed to see a few ($2$--$10$)
additional lensed galaxies. If by any chance the observed lens is
suspected to be at higher redshift, we have to survey a longer strip,
corresponding to $L$ of perhaps a few arcmin (see the bottom panel of
Fig.~\ref{fig:surface_dens}). It is also clear that even a limited
resolution of the survey of 1'' will not significantly change the
strategy (see dashed lines in Fig.~\ref{fig:surface_dens}) since most
of the observed lensed sources will be far from the string and
therefore sufficiently separated in angle.

2) $a=1$: the string exhibits a pure random walk. It is clear that the
total length of the string that we need to encompass will be the same
as in 1) in order to observe several additional lenses. Since the
random walk extends in two directions from the observed lens, we need
to make follow up observations in the circle of radius of
$R=\sqrt{L\xi/2}$ away from the lens (see Fig.~\ref{fig:string}, right
drawing)\footnote{Taking into account the projection of the random walk
on the sky makes the actual required length $L$ larger by $4/\pi$, but
we ignore this small correction.}. This depends on the correlation
length $\xi$. If $\xi > L/2$, the string is straight in the region of
interest and the estimate in 1) holds. So we need only consider $\xi <
L/2$. Therefore the largest diameter we need to cover will be for the case
of $\xi=L/2$, giving $2R=L$ ($\approx 1'$ for the lens candidate of
Ref.~\cite{CSL1}).

It is now clear that the sufficiently effective strategy for any value
of $0\leq a\leq 1$ should be the same as in case 2) above: following
up candidates in the circle of diameter $L$ ($\approx 1'$ for
$z_l\lesssim 0.5$) centered on the lens guarantees finding at least a
few additional lenses, provided that the original lens was caused by
the cosmic string. This statement is quite robust. One can of course
imagine that the string conspires and its direction near the lens
becomes nearly parallel to the line of sight, but the skeptical
astronomer can decrease the odds of this happening by simply following
up a larger area around the observed lens. Note also that lensing by
conventional sources will not cause confusion, since only about 0.1
percent of galaxies are expected to be lensed by large-scale
structure, which implies roughly 1 conventional lens for every 50 
cosmic string lenses in the vicinity of the observed event.

A more serious potential problem is that of false positives --- chance
alignments of galaxies located nearby. It can be seen that this is not
a problem for two reasons. First, the galaxies are on average spread
apart by $6''$, only a small fraction will be closer to each other
than $\sim 2''$, and those can be separated by photometric redshifts,
which are nowadays quite accurate ($\sigma_z\lesssim
0.05$~\cite{Refregier}).  Second, the very few galaxies that
accidentally happen to be nearby both in angle and redshift will in
general have different shapes while, recall, lensing by a cosmic
string does not cause image distortions. Comparing the shapes of
candidate lenses is another way of filtering out false positives. 

The most serious concern are the binary galaxies, and they are the
reason that careful follow up of the lens candidates is preferred.

\section{Conclusions}

Cosmic strings can produce observable lensing signatures even if they are
light enough  to be irrelevant for structure formation in the universe.
Previous reports of objects being lensed by a cosmic string have not 
been confirmed after follow up observations. It remains to be seen what 
further observations will tell us about the current candidate \cite{CSL1}. 

Future wide-field surveys, such as SNAP and LSST, 
will be able to see thousands of lenses caused by a single infinite 
string with linear density $G\mu/c^2\approx 10^{-6}$. However, one cannot 
guarantee that these surveys will either find evidence for strings or else 
significantly improve limits on their abundance and energy density
simply because the number of lenses caused by cosmic strings will 
be dwarfed by a much larger number of galaxies gravitationally 
lensed by large-scale structure in the universe.

Tackling a somewhat different problem, we have argued that confirming
or refuting the hypothesis of lensing by the cosmic string, {\it
given} one reported observation of such an event, is in principle
straightforward. Regardless of what shape the string has, follow up
observations in the circle of diameter $L$ centered on the observed
lens will uncover at least a few galaxies with split images. For a
string located at $z_l\lesssim 0.5$, $L\approx 1'$. Since the original
observation will presumably report the source redshift $z_s$ and
splitting $\Delta\alpha$, using Eq.~(\ref{eq:splitting}) one can bound
the value of the string redshift $z_l$ and, using
Eq.~(\ref{eq:N_obs}), compute the required value of follow up diameter
$L$ so that at least a few additional multiply imaged galaxies are
guaranteed to be seen. Since there will be a total of about 100
galaxies per arcmin$^2$, photometric redshifts have a sufficiently
good accuracy ($\sigma_z\lesssim 0.05$) to select lensing
candidates. Furthermore, lensing by large-scale structure will be
subdominant in this region and will not cause confusion. The only
serious complication is presence of binary systems which will require
more careful follow up.

\vspace{0.3cm}
\begin{acknowledgments}
This work was supported by a Department of Energy grant to the
particle astrophysics theory group at CWRU. DH thanks Chuck Keeton
for useful conversations.
\end{acknowledgments}

\end{document}